\documentclass[aps,prb,twocolumn,superscriptaddress,showpacs,showkeys,a4paper,english]{revtex4-1}
\usepackage{amsmath,epsfig,amssymb,subfigure,bm,dsfont}
\usepackage{times}
\usepackage{graphicx}
\usepackage{float}
\usepackage{color}
\usepackage{subfigure}
\usepackage{epstopdf}
\usepackage[colorlinks=true,linkcolor=blue,citecolor=blue]{hyperref}
\usepackage{hyperref}
\usepackage{soul}
\usepackage{ulem}
\usepackage{bbold}

\usepackage[english]{babel}
\usepackage[OT1]{fontenc}
\sethlcolor{yellow}

\begin{document}

\title{AB-initio optimization principle for the ground states of translational invariant strongly-correlated quantum lattice models}

\author{Shi-Ju~Ran}
\email[Corresponding author. ]{Email: shi-ju.ran@icfo.es}
\affiliation{ICFO-Institut de Ciencies Fotoniques, The Barcelona Institute of Science and Technology, 08860 Castelldefels (Barcelona), Spain}

\begin{abstract}
  In this work, a simple and fundamental numeric scheme dubbed as ab-initio optimization principle (AOP) is proposed for the ground states of translational invariant strongly-correlated quantum lattice models. The idea is to transform a nondeterministic-polynomial-hard ground state simulation with infinite degrees of freedom into a single optimization problem of a local function with finite number of physical and ancillary degrees of freedom. This work contributes mainly in the following aspects: 1) AOP provides a simple and efficient scheme to simulate the ground state by solving a local optimization problem. Its solution contains two kinds of boundary states, one of which play the role of the entanglement bath that mimic the interactions between a supercell and the infinite environment, and the other give the ground state in a tensor network (TN) form. 2) In the sense of TN, a novel decomposition named as tensor ring decomposition (TRD) is proposed to implement AOP. Instead of following the contraction-truncation scheme used by many existing TN-based algorithms, TRD solves the contraction of a uniform TN in an opposite clue by encoding the contraction in a set of self-consistent equations that automatically reconstruct the whole TN, making the simulation simple and unified; 3) AOP inherits and develops the ideas of different well-established methods, including the density matrix renormalization group (DMRG), infinite time-evolving block decimation (iTEBD), network contractor dynamics, density matrix embedding theory, and etc., providing a unified perspective that is previously missing in this fields; 4) AOP as well as TRD gives novel implications to existing TN-based algorithms: a modified iTEBD is suggested and the 2D AOP is argued to be an intrinsic 2D extension of DMRG that is based on infinite projected entangled pair state. This paper is focused on one-dimensional quantum models to present AOP. The benchmark is given on transverse Ising Chain and 2D classical Ising model, showing a remarkable efficiency and accuracy of AOP.
\end{abstract}

\maketitle

\section{Introduction}

Incredible success has been achieved in, e.g. quantum chemistry, condensed matter physics and material sciences, benefiting from high unification and commercialization of density functional theory and the first principle approaches \cite{DFTreview}. But these techniques suffer severe limitations, especially for the quantum many-body systems with strong correlations, which is one of the central but challenging topic in modern physics. For example, the two-dimensional (2D) Heisenberg models with geometrical frustration \cite{Frustration1,Frustration2}, e.g., the kagome anti-ferromagnet \cite{KagomeA,kagomeDMRG}, are believed to realize the exotic quantum spin liquids, which has no symmetry breaking even at zero temperature \cite{QSL} and may exhibit exotic topological orders \cite{TopoOrder}. Hubbard model and its various extended versions \cite{SCHubbard} are promising to provide theoretical explanations for high-temperature superconductivity \cite{HTSC}. Unfortunately, analytical solutions for such models are extremely rare, and numeric approaches became fatally important in this field.

Since the Hilbert space increases exponentially with system size, exact diagonalization can only handle small systems, thus has strong finite size effects. Quantum Monte Carlo, which has an extremely wide applications including calculating ground states, excitations, Green functions and dynamic problems \cite{QMCGreen,QMCrev1,QMCrev2}. However, QMC suffers the notorious ``sign'' problem when calculating frustrated quantum spin models and fermion models away from half-filling, which is shown to be nondeterministic polynomial (NP) hard \cite{QMCsign}. 

In recent years, theories and algorithms based on numeric renormalization group \cite{NRG} and tensor network (TN) representation have been through a rapid development. Density matrix renormalization group \cite{DMRG} is remarkably accurate for one-dimensional (1D) systems. As it is based on a matrix product state (MPS) \cite{MPS} from that is essentially a 1D state representation, DMRG becomes inefficient for large 2D systems. Meanwhile, MPS-based algorithms such as (infinite) time-evolving block decimation (TEBD) was proposed and generalized to 2D quantum systems using an intrinsic 2D state representation named projected entangled pair state (PEPS) \cite{PEPS} (also called tensor product state). PEPS can be regarded as a tensor network (TN) that is defined as contractions of local tensors, and can faithfully represent non-critical 2D quantum states \cite{MPSPEPS}. Then, the simulations mostly become the calculations of TN contractions. The related algorithms \cite{PEPS,iTEBD,ODTNS,TRGsimp,SRG,CTMRG,NCD} have no ``sign'' problem and are able to access infinite systems by using translational invariance. 

Beyond quantum many-body physics, amazing potential of TN is demonstrated also in many other fields. For example, a quantum algorithm was proposed to prepare injective PEPS \cite{PEPS_inj} in a quantum computer \cite{QCP_PEPS1}. The contraction of a TN can be applied to solve fundamental computational tasks, such as search problems \cite{Search}. Such an extreme wide range of TN applications justifies the importance and usefulness of a general and unified numeric scheme of TN.

Normally, a TN-based algorithm follows a contraction-truncation scheme in order to optimally solve targeted physical problems \cite{TNcluster}. The goal of the simulation becomes contracting the corresponding TN, which is usually NP-hard \cite{NPhard}. For example, the time evolution is actually the contraction of a 2D TN formed by the local evolution operators \cite{iTEBD}. The computation of the fidelity between two PEPS's becomes the contraction of the corresponding TN \cite{Fidelity}.

Many approaches have been proposed to deal with TN contractions. One important way is based on tensor renormalization group (TRG) \cite{TRG1,TERG,TRGsimp,SRG,CTMRG}. The general idea is to transform and contract the local tensors in the TN, so that the number of the physical systems of one local tensor increases in a coarse-graining way. In other words, local degrees of freedoms are summed over in a specific order, so that several local tensors are ``coarse-grained'' into one larger tensor which represents a larger physical subsystem, and the total number of the tenors in the TN decreases. In this process, truncations have to be introduced since the bond dimensions of the local tensors increase exponentially as the renormalization goes on. 

Thus, one key ingredient of a TRG-based algorithm is how to truncate optimally, which determines the complexity of coding, computational cost and accuracy of the algorithm. For instance, the simple update algorithm \cite{TRGsimp} provides a local truncation scheme that is extremely efficient and easy to implement, and the bond dimensions of the tensors can be very large. The full update algorithms \cite{PEPS,SRG,CTMRG,FFupdate}, where in principle the whole TN should be contracted to obtain globally optimal truncations, are normally more accurate but expensive. It allows a comparatively smaller bond dimension. Recently, symmetries are considered in the TN algorithms so that a much larger bond dimension becomes tolerable \cite{PEPSU1,Qspace}. However, how to balance between the truncation scheme and the bond dimensions is still under debate.

Meanwhile, the error of a PEPS in the contraction-truncation scheme comes from two aspects: the truncations of the PEPS, and the truncations to obtain the environment of the PEPS. Specifically speaking, to obtain each truncation (or variation) in a full update, one needs to contract a TN where more truncations are inevitable \cite{PEPS,SRG,CTMRG,FFupdate}. It makes the control or estimate of the error a challenging issue. One typical way to obtain the error estimate is to extrapolate the bond dimension to infinite under a reasonable assumption. A recent work shows that a good estimate of the error can be achieved by a polynomial fit against the bond dimension \cite{TNerr}.

Another issue of the contraction-truncation scheme rises from the diversity of the geometries of 2D lattices. Since the TN varies for different lattices, the contraction process can be totally different, and so are the efficiency and computational cost. In other words, the algorithm strongly relies on the details of the models, which hinders further development and applications of TN-based algorithms, especially for the non-specialists. Thus, an efficient, unified and simple scheme that is less dependent on models' details is urgently needed.

Recently, a new clue has been proposed to deal with TN's in the other way around. Instead of thinking about how to contract an infinite TN to a local object with optimal truncations, the key idea is to find a set of local self-consistent equations, from which the TN itself can be automatically reconstructed from such local equations to infinite. In other words, the clue is to \textit{encode} a uniform TN into local equations. This idea can be traced back to the canonicalization of MPS \cite{Canonical}, where a uniform MPS can be reconstructed from the self-consistent canonical conditions. Its two-dimensional generalization was proposed, where the optimal tree approximation of an infinite PEPS is encoded in the super-orthogonal conditions \cite{ODTNS}, which implies a modified version of Tucker decomposition \cite{TD}. Then, the theory of network contractor dynamics (NCD) \cite{NCD} was proposed to generalize the encoding idea of PEPS to any uniform TN, where the self-consistent equations are determined by the rank-1 decomposition \cite{Rank1} of the local tensor. Such schemes largely simplify the calculations of TN's, giving birth to novel concepts and efficient algorithms which have been shown to be greatly useful to, e.g., detecting criticality and calculating ground-state and thermodynamic properties of many-body systems. 

However, one can see that only the TN's with no loops (such as a 1D MPS or a tree TN) have been successfully encoded. For a uniform TN with a regular geometry (e.g. a square or cubic TN), the encoding just gives an optimal ``Bethe approximation'' that contains no loops. It is still unknown how to construct the self-consistent equations that directly encode a uniform 2D TN.

It is worth mentioning that building self-consistent equations is one of the most successful and important ideas in physics, which is fundamental to, e.g., mean-field theories, density functional theories and first principle approaches \cite{DFTreview,DMFTrev}. Recently, density matrix embedding theory \cite{DMET,DMET1} developed this idea to strongly-correlated fermionic models, where an infinite bulk system is self-consistently mapped into an impurity model with an entanglement bath. Later, an extended version of DMET for spin lattice models was proposed based on a product cluster state \cite{DMET2}.

In this work, the ideas mentioned above are extended for better considering the quantum entanglement in many-body systems \cite{EntangleRev}, and a simple and unified scheme dubbed as the \textit{ab-initio optimization principle} (AOP) approach for simulating the ground states of quantum many-body systems with translational invariance is proposed. With a given Hamiltonian $\hat{H}$, a set of self-consistent equations [Eqs. (\ref{eq-eigen2}) - (\ref{eq-eigen1})] are built, which transform the NP-hard ground state simulation \cite{NPhard} problem $|\phi_0\rangle = \min_{|\phi\rangle} \langle \phi| \hat{H}|\phi\rangle$ to an optimization problem of a local function $\mathcal{F}$ [Eq. (\ref{eq-Ffunction})] that contains only finite degrees of freedom. The solution of the optimization has two kinds of boundary states, one of which gives the ground state in the form of TN, and the other play the role of the ``entanglement bath'', providing an optimal approximation of the entanglement between the supercell and the infinite environment.

\begin{figure}[tbp]
	\includegraphics[angle=0,width=0.85\linewidth]{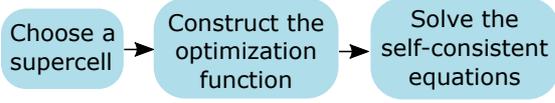}
	\caption{(Color online) Three steps to implement ab-initio optimization principle approach.}
	\label{fig-FOPsteps}
\end{figure}

The way to implement AOP is quite simple and generally independent of the details of the models, where there are three steps (Fig. \ref{fig-FOPsteps}). \textit{Step 1}: choose a proper supercell that is consistent with the translational invariance of the Hamiltonian $\hat{H}$. \textit{Step 2}: construct the operator $\hat{\mathcal{F}}$ (Fig. \ref{fig-Fboundary}) of the supercell from $\hat{H}$. $\hat{\mathcal{F}}$ determines the optimization function $\mathcal{F}$ that is to be maximized. \textit{Step 3}: start with a set of randomly initialized boundary states and solve the optimization problem.  

The robustness of AOP is justified by the TN scheme. From the self-consistent equations, the optimization of $\mathcal{F}$ is equivalent to the global optimization of the zero-temperature density matrix, i.e. $\max_{|\phi\rangle} \lim_{\beta \to \infty} \langle \phi|e^{-\beta \hat{H}} |\phi\rangle$ under the assumption that $|\phi\rangle$ is in an infinite MPS (for 1D $\hat{H}$) or PEPS (for 2D $\hat{H}$) form. The \textit{tensor ring decomposition} (TRD) is proposed [Fig. \ref{fig-TRD} (a)], which amazingly encodes the infinite TN in a local function. When the model is represented in a uniform TN, the only step to calculate its contraction is to decompose the local tensor with TRD. AOP provides a wide connection among well-established methods, including mean-field theory, infinite time-evolving block decimation (iTEBD) \cite{iTEBD}, density matrix renormalization group (DMRG) \cite{DMRG} and network contractor dynamics (NCD) \cite{NCD}. A modified iTEBD and an intrinsic 2D version of DMRG are suggested in AOP. TRD is shown to be closely related to rank-1 decomposition \cite{Rank1} and tensor-train decomposition \cite{TTD} in multi-linear algebra \cite{TDrev}. 

The paper is organized as following. Firstly, by taking a 1D quantum chain as example, I show how to obtain the optimization function $\mathcal{F}$ from $\hat{H}$, where the relation between AOP and the mean-field theory is discussed. Secondly, the construction of the self-consistent equations is shown, where $\mathcal{F}$ is maximized. Then, AOP is discussed in the scheme of TN, where the tensor ring decomposition is proposed and shown to locally encode the infinite TN. An alternating-least-square algorithm is introduced to solve the optimization, which is benchmarked on 1D transverse Ising chain and 2D classical Ising model. The results show that the ground state is accurately obtained at the critical point, indicating that AOP can precisely capture the strong correlations of the quantum many-body system. An equivalence between the supercell size and the dimension cut-off is proposed to physically explain the ``finite size effect'' in AOP. The AOP as well as the TRD for 2D quantum models are presented. Finally, the algorithmic implications of AOP are discussed, and a summary is given. 

\section{Construct the optimization function with boundary states}

Below, I take an infinite quantum chain with nearest-neighbor interactions as the example, whose Hamiltonian reads
\begin{eqnarray}
  \hat{H} = \sum_{i} \hat{H}_{i,i+1}.
  \label{eq-H}
\end{eqnarray}
The optimization function will be constructed in such a way, where the second-order Trotter-Suzuki decomposition of the zero-temperature density matrix of the system $e^{-\beta \hat{H}}$ $(\beta \rightarrow 0)$ is encoded in the self-consistent equations. Note that most of the discussions below can be readily generalized to two or higher dimensions. 

First, choose the supercell which can simply be a finite block with $N$ sites. The operator $\hat{\mathcal{F}}$ that determines the optimization function is formed by two parts: bulk and boundary. Define the bulk Hamiltonian as $\hat{H}^B = J \sum_{i=1}^{N-1} \hat{H}_{i,i+1}$. Minimizing just $\hat{H}^B$, i.e. $\min(\langle \phi | \hat{H}^B | \phi \rangle)$, surely just gives the result of exact diagnalization with $N$ sites, which suffers a strong finite-size effect. The problem is how to introduce a proper boundary to mimic the interactions among the supercells.

\begin{figure}[tbp]
  \includegraphics[angle=0,width=1.0\linewidth]{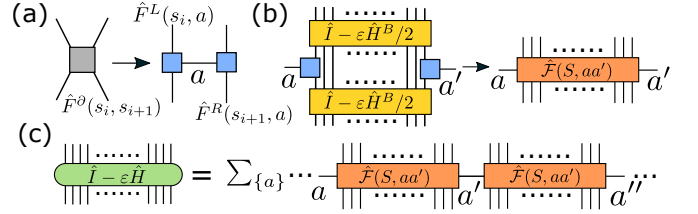}
  \caption{(Color online) (a) The operator $\hat{F}^{\partial}(s_i,s_{i+1})$ is written as a summation of $\hat{F}^L(s_i,a)$ and $\hat{F}^R(s_{i+1},a)$ with eigenvalue decomposition [Eq. (\ref{eq-Fboundary})]. (b) Given by Eq. (\ref{eq-getF}), $\hat{\mathcal{F}}(S,aa')$ with $S = (s_1,\cdots,s_N)$ is obtained by acting $\hat{F}^{R}(s_1,a)$ and $\hat{F}^{L}(s_N,a')$ on the first and last sites of $(\hat{I} - \varepsilon \hat{H}^B/2)$, respectively, with $\hat{H}^B$ the bulk Hamiltonian of a supercell. (c) The relation between the total Hamiltonian $\hat{H}$ and the operator $\hat{\mathcal{F}}(S,aa')$ given by Eq. (\ref{eq-FH}).}
  \label{fig-Fboundary}
\end{figure}

The Hamiltonian on the boundary $\hat{H}^{\partial}(s_N,s_{N+1}) = \hat{H}_{N,N+1}$ is the interaction(s) between two adjacent supercells. Make a shift of it as
\begin{eqnarray}
  \hat{F}^{\partial}(s_N,s_{N+1}) = \hat{I} - \varepsilon \hat{H}^{\partial}(s_N,s_{N+1}),
\end{eqnarray}
with $\varepsilon$ a small number. This shift will not change the ground state. Introduce an ancillary particle $a$ and rewrite $\hat{F}^{\partial}$ as a sum of operators [Fig. \ref{fig-Fboundary} (a)] as
\begin{eqnarray}
  \hat{F}^{\partial}(s_N,s_{N+1}) = \sum_a \hat{F}^{L}(s_N,a) \otimes \hat{F}^{R}(s_{N+1},a).
  \label{eq-Fboundary}
\end{eqnarray}
Eq. (\ref{eq-Fboundary}) can be easily achieved by eigenvalue decomposition. Then the operator $\hat{\mathcal{F}}(S,aa')$, with $S = (s_1,\cdots,s_N)$ representing the $N$ physical particles in the super-cell, is defined as
\begin{eqnarray}
  \hat{\mathcal{F}}(S,aa')= \tilde{H}^B \hat{F}^{R}(s_1,a)^{\dagger} \hat{F}^{L}(s_N,a) \tilde{H}^B,
  \label{eq-getF}
\end{eqnarray}
where one has $\tilde{H}^B = \hat{I} - \varepsilon \hat{H}^B/2$, $\hat{F}^{R}(s_1,a)^{\dagger}$ and $\hat{F}^{L}(s_N,a')$ act on the first and last sites of the super-cell, respectively [Fig. \ref{fig-Fboundary} (b)]. One can see that $\hat{\mathcal{F}}(S,aa')$ contains $N$ physical particles in the supercell and two ancillary particles on the boundaries.  $\hat{\mathcal{F}}(S,aa')$ has a clear relation with $\hat{H}$ [Fig. \ref{fig-Fboundary} (c)] that is
\begin{eqnarray}
  \sum_{aa'a''\cdots} \cdots \hat{\mathcal{F}}(S,aa') \hat{\mathcal{F}}(S',a'a'') \cdots = \hat{I}-\varepsilon \hat{H} + O(\varepsilon^2).
  \label{eq-FH}
\end{eqnarray}
One can see that Eq. (\ref{eq-FH}) gives the second-order Trotter-Suzuki decomposition of $e^{-\varepsilon \hat{H}}$ \cite{Trotter}.

In fact, the ancillary particles $\{a\}$ carry the quantum entanglement between different super-cells. To see this, I take the Heisenberg model as an example, where $\{a\}$ takes from $0$ to $3$. If one limits the number of the states of $a$ in Eq. (\ref{eq-Fboundary}) as one, i.e. $\hat{F}^{\partial} \simeq \hat{F}^{L}(s_i,0) \otimes \hat{F}^{R}(s_{i+1},0)^{\dagger}$, then the operators $\hat{F}^{L}(s_i,0)$ and $\hat{F}^{R} (s_{i+1},0)^{\dagger}$ simply give a mean-field in Eq. (\ref{eq-FH}), i.e. 
\begin{eqnarray}
  \hat{F}^{L(R)}(s_i,0) = \sum_{\alpha} \tilde{h}^{\alpha}_i \hat{s}^{\alpha}_i,
  \label{eq-meanfield}
\end{eqnarray}
with $\hat{s}^{\alpha}_i$ ($\alpha=x,y,z$) the spin operators and $\tilde{h}^{\alpha}_i$ the mean-field on the $i$th site. I shall remark that $\tilde{h}^{\alpha}_i$ is not obtained by simply taking the dominant eigenvectors in Eq. (\ref{eq-Fboundary}), but achieved in a self-consistent way (by Eq. (\ref{eq-Frank1})). In this case, the wave function of the whole system is just the tensor product of the states of super-cells, thus there will be no quantum entanglement among different supercells.

With the operator $\hat{\mathcal{F}}(S,aa')$, introduce the \textit{optimization function} $\mathcal{F}$ [Fig. \ref{fig-Fminimization} (a)] as
\begin{eqnarray}
  \mathcal{F} = \langle L_{a'\mu'\nu'}| \langle A_{S \nu \nu'}| \hat{\mathcal{F}}(S,aa') | A_{S \mu \mu'} \rangle |R_{a\mu \nu}\rangle,
  \label{eq-Ffunction}
\end{eqnarray}
where $\mu, \nu, \mu', \nu'$ that take from $0$ to $\chi-1$ ($\chi$ is a positive integer dubbed as the ring rank) represent the ancillary particles and are traced out in Eq. (\ref{eq-Ffunction}). The \textit{boundary states} $| A_{S\mu \mu'} \rangle$, $|L_{a'\mu' \nu'}\rangle$ and $|R_{a\mu \nu}\rangle$ are three normalized vectors defined in the corresponding ancillary and physical space. I will show below when $\mathcal{F}$ is maximized, the ground state can be given by $| A_{S\mu \mu'} \rangle$ [see Eq. (\ref{eq-GSMPS})].

\section{Ab-initio optimization principle with tensor network scheme}

\begin{figure}[tbp]
	\includegraphics[angle=0,width=0.9\linewidth]{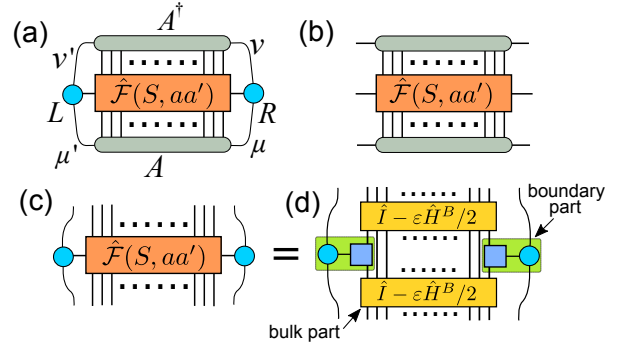}
	\caption{(Color online) Graphic representations of (a) the optimization function $\mathcal{F}$ in Eq. (\ref{eq-Ffunction}) that is to be maximized, (b) the operator $\hat{\mathcal{M}}$ in Eq. (\ref{eq-Mmatrix}), and (c) $\hat{\mathcal{H}}$ in Eq. (\ref{eq-Heffective}) that consists of (d) the bulk and boundary parts. }
	\label{fig-Fminimization}
\end{figure}

The ground state of a 1D quantum many-body system obtained by AOP approach is actually in the form of an MPS. To see this, define two operators [Figs. \ref{fig-Fminimization} (b) and (c)] as
\begin{eqnarray}
 \hat{\mathcal{M}}(aa'\mu \mu' \nu \nu') = \langle A_{S\nu \nu'}| \hat{\mathcal{F}}(S,aa') | A_{S\mu \mu'}\rangle.
 \label{eq-Mmatrix} \\
 \hat{\mathcal{H}}(S\mu \mu' \nu \nu') = \langle L_{a'\mu'\nu'}| \hat{\mathcal{F}}(S,aa') |R_{a\mu \nu}\rangle, 
 \label{eq-Heffective}
\end{eqnarray}
Then, $\mathcal{F}$ is maximized if the following self-consistent equations are fulfilled
\begin{eqnarray}
\langle L_{a'\mu\nu}| \hat{\mathcal{M}}(aa'\mu \mu' \nu \nu') = \mathcal{F}_{max} \langle L_{a'\mu'\nu'}|,
\label{eq-eigen2}\\
\hat{\mathcal{M}}(aa'\mu \mu' \nu \nu') |R_{a\mu \nu}\rangle = \mathcal{F}_{max} |R_{a\mu'\nu'}\rangle, 
\label{eq-eigen3}\\
\hat{\mathcal{H}}(S\mu \mu' \nu \nu') |A_{S\mu \mu'}\rangle \simeq \mathcal{F}_{max} |A_{S\nu \nu'}\rangle,
\label{eq-eigen1}
\end{eqnarray}
with $\mathcal{F}_{max}$ a constant giving the maximum of $\mathcal{F}$. Eqs. (\ref{eq-eigen2}-\ref{eq-eigen1}) mean that $\langle L_{a'\mu\nu}|$ and $|R_{a\mu \nu}\rangle$ are the left and right dominant eigenstates of $\hat{\mathcal{M}}(aa'\mu \mu' \nu \nu')$, respectively. For $|A_{S\mu \mu'}\rangle$, it is the right dominant eigenstate of $\hat{\mathcal{H}}(S\mu \mu' \nu \nu')$ under a constraint, which will be immediately discussed in the following part of this section using the language of TN and MPS. The graphic illustrations of Eqs. (\ref{eq-eigen2}-\ref{eq-eigen1}) and a detailed deduction of these constrained eigenvalue problems are given in Appendix B. graphic Note that  $\langle L_{a'\mu\nu}|$ and $|R_{a\mu \nu}\rangle$ are conjugate to each other if there exists eigenvalue decomposition of $\hat{\mathcal{M}}(aa'\mu \mu' \nu \nu')$. This is usually true in a physical systems, when it possesses the invariance under a spatial mirror reflection, i.e. Eq. (\ref{eq-Mmatrix}) is Hermitian. The existence of the solution for Eq. (\ref{eq-eigen3}) is justified by the fact that the Hamiltonian (as well as the density matrix of the system) is Hermitian.

In the TN language, the operator $\hat{\mathcal{F}}(S,aa')$ and the boundary states $|A_{S\mu \mu'}\rangle$, $\langle L_{a'\mu'\nu'}|$ and $|R_{a\mu \nu}\rangle$ are given by tensors $\mathcal{F}_{ SS' aa'}$, $A_{S\mu \mu'}$, $L_{a'\mu'\nu'}$ and $R_{a\mu \nu}$. One has, for example, $\hat{\mathcal{F}} (S,aa') =\sum_{SS' aa'} \mathcal{F}_{ SS' aa'} |a' \rangle |S'\rangle \langle S | \langle a| $ and $|A_{S\mu \mu'}\rangle = \sum_{S\mu \mu'} A_{S\mu \mu'} |S\rangle |\mu\rangle |\mu'\rangle$, where $|\ast \rangle$ represents the local basis in the corresponding physical or ancillary space. Then the products of operators and vectors in the Hilbert space become contractions of the corresponding tensor indexes. 

\begin{figure}[tbp]
	\includegraphics[angle=0,width=0.9\linewidth]{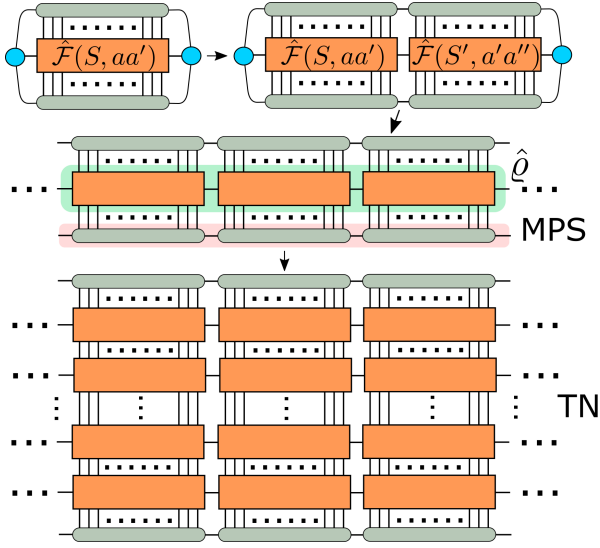}
	\caption{(Color online) From the self-consistent equations [Eqs. (\ref{eq-eigen2}-\ref{eq-eigen1})], the optimization function $\mathcal{F}$ in Eq. (\ref{eq-Ffunction}) can be written as $\mathcal{F} = \langle \Phi | \hat{\varrho} | \Phi \rangle$, where one has $\hat{\varrho} = \hat{I} - \varepsilon \hat{H}$ (green shadow) and $| \Phi \rangle$ is the ground state in an MPS form (red shadow) given by Eq. (\ref{eq-GSMPS}). Then from the eigen equation $\hat{\varrho} | \Phi \rangle = C | \Phi \rangle$, an infinite TN can be constructed.}
	\label{fig-Revursive}
\end{figure}

Using Eqs. (\ref{eq-eigen2}) and (\ref{eq-eigen3}), one can add $\hat{\mathcal{M}}(aa'\mu \mu' \nu \nu')$'s in Eq. (\ref{eq-Ffunction}), as shown in Fig. \ref{fig-Revursive}. This can be repeated for infinite times, after which one gets $\mathcal{F} = \langle \Phi | \hat{\varrho} | \Phi \rangle$, where the operator in the middle (green shadow) is actually $\hat{\varrho}=(\hat{I} - \varepsilon \hat{H})$ in Eq. (\ref{eq-FH}), and the state $|\Phi \rangle$ has an MPS form \cite{MPS} (red shadow) that reads
\begin{eqnarray}
|\Phi \rangle = \sum_{\{S\}} \sum_{\{\mu\}} \cdots A_{S\mu \mu'} A_{S'\mu' \mu''} \cdots |\{S\}\rangle,
\label{eq-GSMPS}
\end{eqnarray}
with $\{S\} = (\cdots,S,S',\cdots)$. Because $|\Phi \rangle$ maximize $\langle \Phi | \hat{\varrho} | \Phi \rangle$, such an MPS optimally gives the dominant eigenstate of $\hat{\varrho}$, which actually is the ground state of $\hat{H}$. 

The tensor $A$ that gives the ground state MPS is the ground state of $\hat{\mathcal{H}}$, which can be regarded as an ``impurity'' model \cite{DMET,DMET1,DMET2}. To be more specific, $\hat{\mathcal{H}}$ consists of two parts: bulk and boundary. The bulk part is the shift of the bulk Hamiltonian, and the boundary part is formed by $\hat{F}^L$, $\hat{F}^R$ [Eq. (\ref{eq-Fboundary})] and the boundary states $\langle L|$ and $|R\rangle$, as shown by the green shadow in Fig. \ref{fig-Fminimization} (d).

Taking one step further, one can repeat for $K \to \infty$ times using the relation $\hat{\varrho} | \Phi \rangle = C | \Phi \rangle$ with $C$ a constant and reconstruct an infinite 2D TN that is formed by the local tensor $\mathcal{F}_{ SS' aa'}$ (Fig. \ref{fig-Revursive}). This TN gives the Trotter-Suzuki decomposition of $e^{-K\varepsilon \hat{H}}$. It means that maximizing $\mathcal{F}$ in Eq. (\ref{eq-Ffunction}) realizes the global optimal contraction of the 2D TN with the MPS. 

AOP grows the infinite TN from a local function, which is different from existing contraction \& truncation schemes. Taking Levin and Nave's TRG \cite{TRG1} as an example, each time after renormalization, the number of tensors is reduced to half of it. But in AOP by following the arrows in Fig. \ref{fig-Revursive}, the number of tensors grows from one to infinite during the reconstruction of the TN.

\begin{figure}[tbp]
	\includegraphics[angle=0,width=1\linewidth]{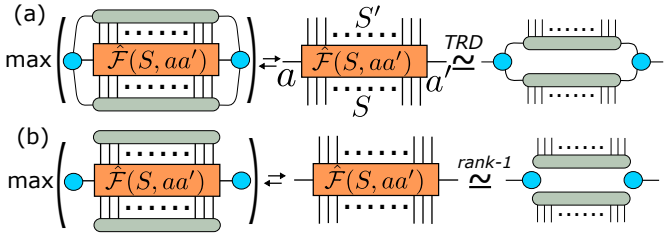}
	\caption{(Color online) Graphic representations of (a) the tensor ring decomposition (TRD) given by Eq. (\ref{eq-TRD}) and (b) the rank-1 decomposition \cite{NCD,Rank1} by Eq. (\ref{eq-Rank1Decomp}). The rank-1 decomposition is special TRD with the ring rank $\chi=1$.}
	\label{fig-TRD}
\end{figure}

What is amazing is that the maximization of $\mathcal{F}$ only corresponds to a specific decomposition of the local tensor $\mathcal{F}_{ SS' aa'}$ [Fig. \ref{fig-TRD} (a)], which is dubbed as \textit{tensor ring decomposition} (TRD)
\begin{eqnarray}
\mathcal{F}_{ SS' aa'} \simeq \mathcal{F}_{max} \sum_{\mu \mu' \nu \nu'=1}^{\chi} A_{S\mu \mu'} L_{a'\mu'\nu'} A_{S' \nu \nu'} R_{a\mu \nu},
\label{eq-TRD}
\end{eqnarray}
where $|\mathcal{F}_{ SS' aa'} - \mathcal{F}_{max} \sum_{\mu \mu' \nu \nu'} A_{S\mu \mu'} L_{a'\mu'\nu'} A_{S' \nu \nu'} R_{a\mu \nu}|$ is minimized and $\chi$ is the ring rank as mentioned above. The TN reconstruction shown above indicates that TRD is an intrinsic higher-order tensor decomposition which encodes the global contraction of the infinite TN.

The ancillary particles $\mu$, $\mu'$, $\nu$ and $\nu'$ in $\mathcal{\hat{H}}$ play the role of carrying the entanglement between a supercell and its infinite environment. In other words, the boundary states $\langle L_{a'\mu'\nu'}|$ and $|R_{a\mu \nu}\rangle$ provide an ``entanglement bath'', which is shown to be critically important to reduce the boundary effect cause by the finiteness of the supercell. 

To see this more clearly, one takes the dimension of the ancillary particles as $\chi=1$. Then, Eq. (\ref{eq-TRD}) becomes
\begin{eqnarray}
\mathcal{F} = \langle L_{a'}| \langle A_{S}| \hat{\mathcal{F}}(S,aa') | A_{S} \rangle |R_{a}\rangle.
\label{eq-Frank1}
\end{eqnarray}
The maximization of Eq. (\ref{eq-Frank1}) gives the so-called rank-1 decomposition \cite{Rank1} of $\mathcal{F}_{ SS' aa'}$ [Fig. \ref{fig-TRD} (b)] that reads
\begin{eqnarray}
\mathcal{F}_{ SS' aa'} \simeq \tilde{\mathcal{F}} A_{S} L_{a'} A_{S'}^{\ast} R_{a},
\label{eq-Rank1Decomp}
\end{eqnarray}
with $\tilde{\mathcal{F}}$ a constant. In this case, The ground state is the tensor product of infinite number of $| A_{S} \rangle$'s, and $\langle L_{a'}|$ and $|R_{a}\rangle$ determine the mean-field in Eq. (\ref{eq-meanfield}). Actually, Eq. (\ref{eq-Rank1Decomp}) leads to the network contractor dynamics (NCD) \cite{NCD}. One can see that the supercell corresponds to the unit cell in NCD. To increase the accuracy, a renormalization procedure is utilized in NCD to increase the unit cell size to infinite. Here in AOP, one increases the ring rank $\chi$ instead, which is more simple and unified. Comparing the physical pictures in AOP and NCD, one can see that the self-consistent equations in NCD (which can be obtained with those in AOP by taking $\chi=1$) result in the optimal tree TN approximation with no loops, while those of AOP lead to the infinite TN with all loops remaining intact. One can also see that rank-1 decomposition is just a special TRD with $\chi=1$. 

\section{Alternating-least-square algorithm with a fixed ring rank}

From the self-consistent equations Eqs. (\ref{eq-eigen2}-\ref{eq-eigen1}), an alternating-least-square algorithm is proposed to efficiently solved the maximization of $\mathcal{F}$. Starting from a randomly initialized $L_{a'\mu\nu}$ and $R_{a\mu \nu}$ with a chosen $\chi$, one calculates $\hat{\mathcal{H}}(S\mu \mu' \nu \nu')$ as well as its dominant eigenstate that is $A_{S\mu \mu'}$; then with the newly obtained $A_{S\mu \mu'}$, calculate $\hat{\mathcal{M}}(aa'\mu \mu' \nu \nu')$ and update $L_{a'\mu\nu}$ and $R_{a\mu \nu}$ with the left and right dominant eigenstate of $\hat{\mathcal{M}}(aa'\mu \mu' \nu \nu')$. Repeat this procedure until the preset convergence is reached.

But if one simply uses the iteration procedure above, the result may converge to a trivial fixed point given by Eq. (\ref{eq-Rank1Decomp}) with the ring rank $\chi =1$. This will bring instability to the calculations. One way to stabilize the non-trivial fixed points with a preset $\chi$ is to constrain $L_{a'\mu\nu}$ and $R_{a\mu \nu}$ to be rank-$\chi$. In detail, each time with updated $L_{a'\mu\nu}$ and $R_{a\mu \nu}$, one calculates $\tilde{L}_{a'\mu\nu}$ and $\tilde{R}_{a\mu \nu}$ that fulfill the following optimizations
\begin{eqnarray}
  &&\max_{\tilde{L}_{a'\mu\nu}} \sum_{a'\mu\nu} \tilde{L}_{a'\mu\nu}^{\ast} L_{a'\mu\nu}, \ while \ \sum_{a'\mu} \tilde{L}_{a'\mu\nu}^{\ast} \tilde{L}_{a'\mu\nu'} = I_{\nu \nu'},\nonumber \\
  &&\max_{\tilde{R}_{a\mu\nu}} \sum_{a\mu\nu} \tilde{R}_{a'\mu\nu}^{\ast} R_{a'\mu\nu}, \ while \ \sum_{a\mu} \tilde{R}_{a\mu\nu}^{\ast} \tilde{R}_{a\mu\nu'} = I_{\nu \nu'},
  \label{eq-isometry}
\end{eqnarray}
with $I_{\nu \nu'}$ a $(\chi \times \chi)$ identity and $\ast$ means conjugate. It means $\tilde{L}_{a'\mu\nu}$ and $\tilde{R}_{a\mu \nu}$ are the optimal isometries that maximize Eq. (\ref{eq-Ffunction}). $\tilde{L}_{a'\mu\nu}$ and $\tilde{R}_{a\mu \nu}$ can be easily obtained by singular value decomposition of $L$ and $R$ \cite{MERAprb}, i.e. $L = U S V^{\dagger}$, $\tilde{L} = U V^{\dagger}$ and similarly for $\tilde{R}$. In this way, one stabilize a non-trivial fixed point with the preset $\chi$. 

Amazingly, important physical information of the system can be extracted from $L$ and $R$: the singular value spectrum $S$ is exactly the (bipartite) entanglement spectrum; the matrix defined as $X=USU^{\dagger}$ gives the dominant eigenstate of the transfer matrix of $\langle \Phi |\Phi \rangle$ (see Appendix B). Note that $X$ is essentially related the the observables (see Appendix C). 

Furthermore, such a way of stabilizing is the result of two constraints in the optimization. In Fig. \ref{fig-Revursive} following the first two arrows, we use Eqs. (\ref{eq-eigen2}) and (\ref{eq-eigen3}) to extend the local contraction to $\langle \Phi | \hat{\varrho} | \Phi \rangle$ with $| \Phi \rangle$ an infinite MPS formed by tensor $A$. To do so, one should fulfill the constraint
\begin{eqnarray}
  \sum_{a\mu\nu} L_{a\mu\nu} L_{a\mu\nu}^{\ast} = \sum_{a\mu\nu} R_{a\mu\nu} R_{a\mu\nu}^{\ast} =1.
\label{eq-constraint1}
\end{eqnarray}
Here, we assume there exists the eigenvalue decomposition of $\hat{\mathcal{M}}$ in Eq. (\ref{eq-Mmatrix}), so that its left and right dominant eigenstates are conjugate to each other. Such a constraint can be fulfilled by directly solving Eqs. (\ref{eq-eigen2}) and (\ref{eq-eigen3}). Then following the third arrow in Fig. \ref{fig-Revursive}, we obtain the whole TN by using the fact that $\langle \Phi | \hat{\varrho} | \Phi \rangle$ is minimized. There is another constraint here, which is the normalization of the MPS
\begin{eqnarray}
\langle \Phi | \Phi \rangle =1.
\label{eq-constraint2}
\end{eqnarray}
The second constraint leads to $\tilde{L}_{a'\mu\nu}$ and $\tilde{R}_{a\mu \nu}$ in Eq. (\ref{eq-isometry}) by solving a generalized eigenvalue equation. The deduction in detail can be found in Appendix B.

With this algorithm, there is only one step to optimally solve the contraction problem of a uniform TN: decompose the local tensor with TRD.

The AOP algorithm is essentially a variational method. In AOP, it is the average of the density matrix that is maximized ($\langle \Phi | \hat{\varrho} | \Phi \rangle$). One can look at the two main steps of the reconstruction of the infinite TN. The first step is from the local optimization function [Eq. (\ref{eq-Ffunction})] to $\langle \Phi | \hat{\varrho} | \Phi \rangle$. The purpose of this step is to let the density matrix and the ground state appear in the formula. The second step is from $\langle \Phi | \hat{\varrho} | \Phi \rangle$ to an infinite 2D TN. The purpose is to guarantee $\langle \Phi | \hat{\varrho} | \Phi \rangle$ is maximized, i.e. the energy is minimized, under the constraint that $| \Phi \rangle$ is normalized. In this sense, AOP is a variational method, but the most significant difference is that AOP translates the variational problem into a set of local eigenvalue equations. To find the optimal solution of the variation is to solve the eigenvalue equations. In this way, AOP makes calculations simple and unified. 

For the computational cost of each iteration on a 1D quantum system, it is about $O(2d^{N} \chi^2 + 2d^{N+4} \chi^2)$ for updating $A_{S\mu \mu'}$, where the first terms is from contracting twice the sparse matrix $\hat{H}^B$ and the second term is from contracting the $\hat{F}^R(s_1,a)$ and $\hat{F}^L(s_N,a')$ ($d$ is the dimension of the physical Hilbert space on each local site). The computational cost for updating $L_{a'\mu\nu}$ and $R_{a\mu \nu}$ is about $O(d^N\chi^2 + d^{N+2}\chi^4 + d^8\chi^4 + d^4 \chi^4)$ with a proper contraction order. Meanwhile, the efficiency of the algorithm is very high, which only takes $O(10^2)$ iterations to reach the convergence (e.g. of the energy) to $10^{-10}$. 

\section{Benchmark}

First, the performance of AOP is tested on the infinite transverse Ising chain. The Hamiltonian is written as
\begin{eqnarray}
  \hat{H} = \sum_{i} \hat{s}^x_i \hat{s}^x_{i+1} - h \sum_{i} \hat{s}^z_i,
  \label{eq-IsingH}
\end{eqnarray}
with $h$ the magnetic field. This model was exactly solved by fermionization \cite{QIsingExact}. At $h=0.5$, a quantum phase transition occurs, where the energy gap vanishes and the quantum entanglement entropy scales logarithmically with the subsystem size \cite{CFM_Ent}. 

\begin{figure}[tbp]
	\includegraphics[angle=0,width=1.0\linewidth]{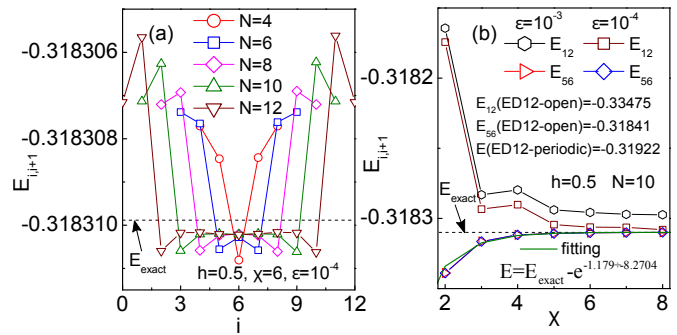}
	\caption{(Color online) (a) The bond energy $E_{i,i+1} = \langle \hat{H}_{i,i+1} \rangle$ versus position $i$ at $h=0.5$, $\chi=6$ and $\varepsilon=10^{-4}$ for different sizes of the supercell $N$. $E_{exact}$ is the exact solution of the infinite chain by fermionization \cite{QIsingExact}. $E_{0,1}$ (and $E_{N,N+1}$) is the bond energy between two adjacent supercells. In the middle of the chain, $E_{i,i+1}$ converges to $E_{exact}$ as $N$ increases, where the error is about $O(10^{-7})$. (b) For $N=10$ and $h=0.5$, $E_{1,2}$ (on the boundary of the supercell) and $E_{5,6}$ (in the middle) versus $\chi$ at $\varepsilon=10^{-3}$ and $10^{-4}$. One can see that the boundary effect decay with $\chi$, where $E_{5,6}$ converges accurately to $E_{exact}$. The systematic error of $E_{1,2}$ is caused by the Trotter-Suzuki error $\sim O(\varepsilon^2)$. By fitting at $\varepsilon = 10^{-4}$, an exponential convergence $E_{5,6} = E_{exact} - e^{-1.179 \chi -8.2704}$ is found. For comparison, $E_{1,2}$ and $E_{5,6}$ obtained by exact diagonalization (ED) at $N=12$ with open and periodic boundary conditions are also shown.}
	\label{fig-Elength}
\end{figure}

To investigate the boundary effect caused by the finiteness of the supercell, the bond energy $E_{i,i+1} = \langle \hat{H}_{i,i+1} \rangle$ versus the position $i$ at the critical point $h=0.5$ for different sizes of the supercell $N$ are calculated and shown in Fig. \ref{fig-Elength} (a). $E_{0,1}$ (and $E_{N,N+1}$) stands for the bond energy between two adjacent supercells. I take $\chi=6$ and $\varepsilon=10^{-4}$. All bond energies are remarkably accurate by comparing with $E_{exact}$ obtained by exact solution on the infinite chain, while in the middle of the chain, $E_{i,i+1}$ converges greatly to $E_{exact}$ as $N$ increase. The error is $O(10^{-6}) \sim O(10^{-7})$. In Fig. \ref{fig-Elength} (b), the bond energies $E_{1,2}$ (on the boundary of the supercell) and $E_{5,6}$ (in the middle) versus $\chi$ at $\varepsilon=10^{-3}$ and $10^{-4}$ are shown with $N=10$ and  $h=0.5$. One can see that as $\chi$ increases, both $E_{1,2}$ and $E_{5,6}$ converge to $E_{exact}$. $E_{1,2}$ suffers a systematic error caused by the Trotter-Suzuki decomposition, which decreases with $\varepsilon$. An exponential convergence is found, e.g., for $\varepsilon = 10^{-4}$, one has $E_{5,6} = E_{exact} - e^{-1.179 \chi -8.2704}$.

For comparison, $E_{1,2}$ and $E_{5,6}$ given by the exact diagonalization (ED) on a $N=12$ chain with open and periodic boundary conditions are shown. The finite size effect of ED is strong. Especially on the boundary, the error is $\sim 10^{-2}$. With AOP, the error, even on the boundary, is only $\sim O(\varepsilon)$, which is around $10^{-4} \sim 10^{-6}$. It suggests that the boundary states provide a good approximation of the interactions between a supercell and its infinite environment.

\begin{figure}[tbp]
	\includegraphics[angle=0,width=1\linewidth]{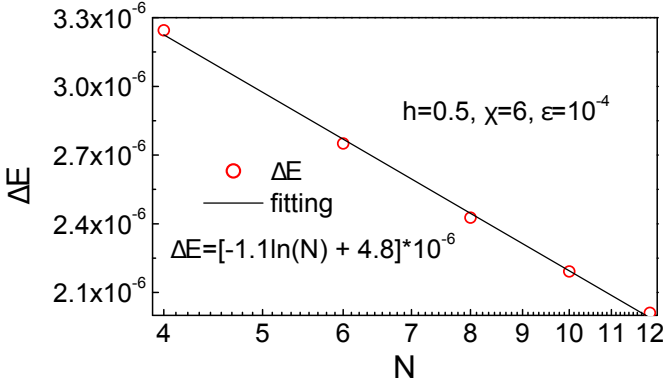}
	\caption{(Color online) The relative error of the average energy per site given by Eq. (\ref{eq-dEav}). By fitting, it is found that $\Delta E$ satisfies a logarithmic relation with $N$ [Eq. (\ref{eq-Errfit})].}
	\label{fig-dEav}
\end{figure}

Fig. \ref{fig-dEav} shows the error of the average energy per site
\begin{equation}
  \Delta E = |\frac{1}{N} \sum_{i=0}^{N-1} E_{i,i+1} -E_{exact}| / |E_{exact}|.
  \label{eq-dEav}
\end{equation}
By fitting, it is found that as $N$ increases, the error $\Delta E$ decreases in a logarithmic way
\begin{equation}
\Delta E = (-1.1\ln N + 4.8) \times 10^{-6}.
\label{eq-Errfit}
\end{equation}
It is amazing that even with $\chi$ fixed, the error still decreases logarithmically, which will be later explained by the entanglement.

\begin{figure}[tbp]
	\includegraphics[angle=0,width=1\linewidth]{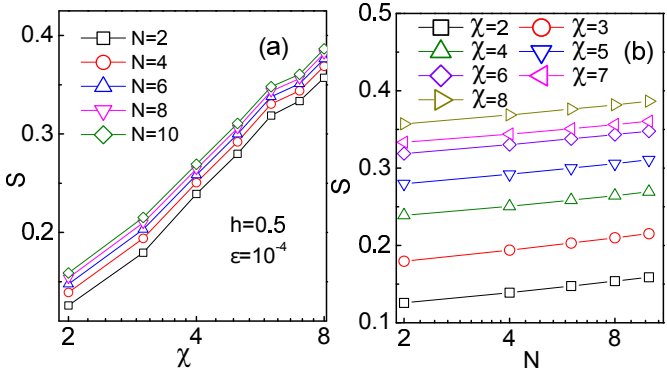}
	\caption{(Color online) The entanglement entropy $S$ versus (a) the ring rank $\chi$ with different supercell size $N$ and (b) versus $N$ with different $\chi$. One can see that $S$ scales linearly with both $\log_2 \chi$ and $\log_2 N$, consistent with the conformal field theory and the former numeric results \cite{CFM_Ent,EntCritic,EntMPS1,EntMPS2,EntMPS3}.}
	\label{fig-EntChi}
\end{figure}

It is known that one of the most important signatures of quantum many-body systems is entanglement \cite{EntangleRev}. To see if AOP can truly capture the many-body characteristics of the system with an ``entanglement bath'', the entanglement entropy at the critical point $h=0.5$ is calculated. The definition of entanglement entropy is written as
\begin{eqnarray}
  S= -\sum_{\mu=1}^{\chi} \lambda_{\mu}^2 \ln(\lambda_{\mu}^2),
\end{eqnarray}
with $\lambda$ the entanglement spectrum between two infinite halves of the chain. Here, the chain is cut at the boundary of the supercells. It is known that at the critical point, $S$ scales logarithmically with both the size of the subsystem and the dimension cut-off of the MPS \cite{CFM_Ent,EntCritic,EntMPS1,EntMPS2,EntMPS3}. Figs. \ref{fig-EntChi} (a) and (b) shows the $\chi$- and $N$-dependence of $S$, respectively. A logarithmic scaling behavior of $S$ versus $\chi$ and $N$ is obtained, consistent with existing results. It suggests that the entanglement is accurately captured by the boundary states with AOP. Such a behavior of $S$ also explains the scaling of the error shown in Fig. \ref{fig-dEav}.

Recall that $N$ (size of the supercell) is not the size of the subsystem when calculating the (bipartite) entanglement. In fact, the subsystem is one half of the infinite chain, thus its size is infinite. The logarithmic scaling versus $N$ implies an equivalence between $N$ and $\chi$. Specifically speaking, the increase of $\chi$ directly enables the state to carry more entanglement, while the increase of $N$, which intuitively reduces the ``finite size effect'' of the supercell, strengthens the entanglement that a fixed $\chi$ can carry. Thus the larger $N$ or $\chi$ is, the more the entanglement can be captured, and the smaller the error would be. In the limit of $N \to \infty$, one can have the exact result (with open boundary condition, precisely speaking) with $\chi=0$. The argument above actually gives the physical picture of the ``finite size effect'' in AOP.

Note that the error of energy away from the critical point is $O(10^{-10})$ with even a smaller number of iteration time to reach the same convergence.

To demonstrate its performance on contracting a 2D TN, the TRD is used to calculate classical 2D Ising model on square lattice at the critical temperature. The obtained free energy is compared with the exact solution by Bethe ansatz \cite{Ising2DExact}.

\begin{figure}[tbp]
	\includegraphics[angle=0,width=1\linewidth]{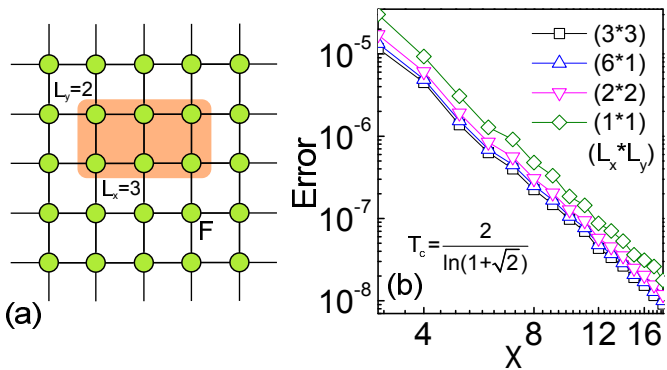}
	\caption{(Color online) (a) The TN of the partition function of 2D Ising model on square lattice is formed by one inequivalent local tensor $F$ given in Eq. (\ref{eq-IsingT}). The yellow shadow shows a tensor cluster with $L_x=3$ and $L_y=2$. (b) For different ($L_x,L_y$), the log-log plot of the error of free energy (per site) versus $\chi$ is shown at the critical temperature $T_c=2/\ln(1+\sqrt{2})$, indicating an algebraic scaling between the error and $\chi$.}
	\label{fig-Ising}
\end{figure}

As shown in Fig. \ref{fig-Ising} (a), the TN of the partition function of 2D Ising model is formed by one inequivalent local tensor $F$ defined as 
\begin{eqnarray}
  F_{s_us_ls_ds_r} = e^{-(s_us_l+s_ls_d+s_ds_r+s_rs_u )/\mathrm{T}},
  \label{eq-IsingT}
\end{eqnarray}
where $s_u$, $s_l$, $s_d$ and $s_r$ are the four spins located on four corners of a square, and $T$ is temperature. By using TRD on tensor clusters with different ($L_x,L_y$), the $\chi$-dependence of the error of free energy (per site) at the critical temperature $T_c=2/\ln(1+\sqrt{2})$ is shown in Fig. \ref{fig-Ising} (b). One can see that the larger $L_x$ or $L_y$ is, the better the precision will be. Different from the exponential scaling at the critical point for 1D quantum model, the log-log plot of the error versus $\chi$ indicates an algebraic scaling in critical 2D classical systems.

Tab. \ref{tab-Ising} shows the errors with iTEBD \cite{Canonical}, NCD \cite{NCD} and AOP by fixing $\chi=16$. Note that in all these three algorithms, $\chi$ stands for the dimension of the MPS. For iTEBD, canonicalization of MPS is employed to reach the optimal truncations of the MPS. For NCD, the cell tensor that is decomposed by rank-1 decomposition is renormalized so that its size reaches infinite. For AOP, I choose tensor clusters with $(L_x,L_y)=(1,1)$ and $(3,3)$ to do the TRD. One can see that AOP bears the best accuracy. A convergence of $O(10^{-12})$ is reached after $O(10)$ times of iteration, implying a great efficiency.

\begin{table}[tbp]
	\caption{Errors of free energy (per site) of 2D Ising model with different methods at the critical temperature. Here, I fix $\chi=16$. For iTEBD, canonicalization of MPS is employed. For NCD, renormalization is used to increase the cell tensor size to infinite. For AOP, I choose $(L_x,L_y)=(1,1)$ and $(3,3)$.}
	\begin{tabular*}{8cm}{@{\extracolsep{\fill}}lcccc}
		\hline\hline
		& iTEBD \cite{Canonical} & NCD \cite{NCD} & AOP (1,1) & AOP (3,3)  \\ \hline
		Error & $1.1 \times 10^{-7}$ & $3.7 \times 10^{-8}$ & $3.1\times 10^{-8}$ & $1.5 \times 10^{-8}$ 
		\\ \hline\hline
		\label{tab-Ising}
	\end{tabular*}
\end{table}

\section{Ab-initio optimization principle for two-dimensional quantum systems} 

Considering that AOP in 1D quantum systems actually realizes the global contraction of a 2D TN (Fig. \ref{fig-Revursive}), it can be used directly to contract 2D TN's regardless of their physical meanings. For this reason, AOP can be used to calculate the observations and the optimal truncations of a PEPS in real/imaginary time evolutions, which are essentially TN contraction problems. Note that since the whole TN is encoded, such a truncation scheme actually realizes the full update \cite{PEPS,SRG,CTMRG} of the PEPS, even with a finite supercell.

Except using its TN contraction scheme to observe or evolve, there is an intrinsic AOP approach for 2D (or higher-dimensional) quantum models, benefiting from the fact that most of the discussions given above are independent of dimensionality. It corresponds to the TRD that encodes a 3D uniform TN. There are also three steps: choose a supercell, construct the optimization function $\mathcal{F}$, and solve the self-consistent equations.

\begin{figure}[tbp]
	\includegraphics[angle=0,width=1\linewidth]{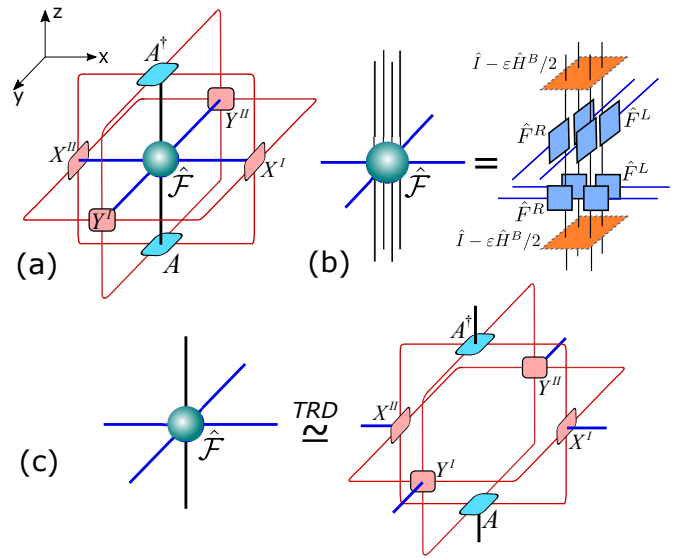}
	\caption{(Color online) (a) Sketch of the optimization function $\mathcal{F}$ for a 2D quantum system, where the operator $\hat{\mathcal{F}}$ is located in the center. The $x$ and $y$ denote the two spatial dimensions of the 2D model. The black bonds in $z$-direction represent the physical indexes of the operator $\hat{\mathcal{F}}$, and the ground state is a PEPS formed by the blue tensor $A$. The four red tensors $X^I$, $X^{II}$, $Y^I$ and $Y^{II}$ are the boundary tensors with ancillary indexes that play the role of ``entanglement bath''. (b) The operator $\hat{\mathcal{F}}$ of a 2D quantum system on square lattice is constructed by choosing a square as the supercell. (c) The sketch of tensor ring decomposition (TRD) of a 3D TN, where the optimization function $\mathcal{F}$ is maximized.}
	\label{fig-TRD_3D}
\end{figure}

Fig. \ref{fig-TRD_3D} (a) gives the construction of the optimization function $\mathcal{F}$ for a 2D quantum system, which is formed by an operator $\hat{\mathcal{F}}$ in the center surrounded by five boundary tensors. The tensors in red ($X^I$, $X^{II}$, $Y^I$ and $Y^{II}$) provide the ``entanglement bath'' that mimic the interactions between a supercell and its environment. Fig. \ref{fig-TRD_3D} (b) shows the sketch of the operator $\hat{\mathcal{F}}$ of 2D quantum model on square lattice, where the supercell is chosen as a square. The maximization of $\mathcal{F}$ leads to the TRD for a 3D TN shown in Fig. \ref{fig-TRD_3D} (c). One can then define three operators and obtain five self-consistent equations in a way similar to Eqs. (\ref{eq-Heffective}) - (\ref{eq-Mmatrix}) and Eqs. (\ref{eq-eigen2}) - (\ref{eq-eigen1}), respectively. For example, $\hat{\mathcal{H}}$ [similar to Eq. (\ref{eq-Heffective})] is defined by taking away $|A\rangle$ and $\langle A|$ from $\mathcal{F}$. Repeatedly using the self-consistent equations in a similar way as in 1D quantum systems, one can readily see that TRD encodes an infinite 3D TN, i.e. the 3D TN can be reconstructed. Meanwhile, the ground state is in the form of a PEPS formed by the tensor in blue $A$. Considering the existence of the solution to the self-consistent equations, it is similar to the AOP for 1D quantum systems: it requires the 2D system to be invariant under spatial mirror reflections, which is usually true for physical models.

The TRD (AOP) for a 3D TN (2D quantum systems) self-consistently contains the TRD for a 2D TN (1D quantum system). Specifically speaking, the optimization function $\mathcal{F}$ for a 3D TN in Fig. \ref{fig-TRD_3D} (a) is a contraction of a 3D tensor cluster. After contracting the black bonds in z-direction, it becomes a 2D object that has the same form as that in Fig. \ref{fig-TRD} (a). The boundary tensors satisfy the consistent equations of the TRD in a 2D TN. For the computational cost by taking square lattice as an example, the leading terms is from the TRD of such a 2D TN, which scales as $O(\chi^{8}d^{2L_x+2L_y})$, where $\chi$ is the ring rank, $L_x$ and $L_y$ give the size of the supercell, and $d$ is the dimension of a physical particle. 

\section{Algorithmic implications of ab-initio optimization principle}

AOP brings useful and novel implications to iTEBD \cite{iTEBD} and DMRG \cite{DMRG,iDMRG}. In iTEBD, one usually follows the evolution-truncation procedure. Specifically speaking, each time after evolving the MPS, the bond dimension increases. Then, one finds the optimal isometry to truncate the bond dimension so that it is limited to a preset cut-off. AOP implies a more efficient way to do this by showing that the local tensor $A_{S\mu \mu'}$ is actually the dominant eigenstate of a matrix formed by the local projector and the isometries [Eq. (\ref{eq-Heffective})]. In detail, with a given $A_{S\mu \mu'}$, one calculates the optimal isometries $L_{a'\mu\nu}$ and $R_{a\mu \nu}$ using iTEBD. Then, one updates $A_{S\mu \mu'}$ by solving the eigen problem given by Eq. (\ref{eq-eigen1}). At the second step, in the language of iTEBD, one evolves the MPS for infinite times till it converges without renewing the isometries. 

In this sense, AOP (in 1D) is similar to iTEBD with a super block of $N$ sites \cite{iTEBD}. Normally, one only takes $N=2$ in iTEBD. One essential difference between these two schemes is that AOP avoids the imaginary time evolution-truncation of the MPS, and unified everything to a more efficient and simple local optimization problem. Note that the high complexities of the existing TN-based algorithms originates mostly from the evolution and truncation tricks, which are strongly relies on the details of the model. AOP largely simplified such complexities. 

Considering that iTEBD is essentially a power algorithm where the projector $e^{-\varepsilon H}$ is plainly acted on a give state to drive it to the ground state, it is not an efficient eigenvalue problem solver. For a given $\varepsilon$ (with a Trotter-Suzuki error $O(\varepsilon^2)$), the computational cost is approximately $\sim 1/\varepsilon$. In AOP where the matrices that are to be explicitly dealt with just have small dimensions [Eqs. (\ref{eq-Heffective}) and (\ref{eq-Mmatrix})], one can use mature techniques (such as the Lanczos algorithm \cite{Lanczos}) to efficiently solve the eigenvalue problems. Consequently, the costs with different $\varepsilon$ are approximately the same. That is one reason for AOP to have a higher efficiency. 

Meanwhile, the tensors $L_{a'\mu'\nu'}$ (and also $R_{a\mu \nu}$) in AOP actually gives the left (right) dominant eigenstate in an MPS that extends in the vertical direction of the TN, which have been proved to be useful in calculating time evolutions \cite{TimeMPS}. 

Besides its high efficiency, it should be emphasized that AOP has a more elegant and natural generalization in 2D quantum systems. Note that the complexity of the evolution-truncation scheme in 2D largely relies on the geometry of the lattice, while in AOP, one avoids such a evolution-truncation scheme and just handles a single optimization problem, which brings great simplification and unification to the simulation in 2D.

As to DMRG \cite{DMRG}£¬ one can access to the infinite chain by utilizing the MPO representation of the 1D Hamiltonian \cite{iDMRG,MPSDMRG}. The boundary tensors $L_{a'\mu\nu}$ and $R_{a\mu \nu}$ is analog to the reduced matrices of the left and right environments. Eq. (\ref{eq-Heffective}) corresponds to the effective Hamiltonian in DMRG. One key difference between (1D) AOP and DMRG is how to obtain the boundary tensors as well as how to reconstruct the translational invariant MPS from the local tensor, which is essential to the algorithms. Besides, in 2D DMRG \cite{DMRG2D}, one still transfers the system into a chain with long-range interactions, and the ground state is in an 1D MPS form that violates the area law of 2D quantum states \cite{AreaLaw}. In this sense, the 2D AOP can be treated as an intrinsic 2D version of infinite DMRG. The key is to properly introduce certain ancillary particles in the optimization problem that leads to a uniform PEPS.

As to TRD,  what it can provide is far more than rank-1 decomposition. One can see that TRD is similar to the so-called tensor train decomposition (TTD) \cite{TTD}, but the boundaries, algorithms and properties are essentially different with each other. Specifically speaking, TTD decomposes a tensor into an MPS with an open boundary, meaning the first and last tensors in the MPS do not directly share any indexes. With TRD, the tensor is decomposed into an periodic MPS formed by the boundary tensors [Figs. \ref{fig-TRD} (a) and \ref{fig-TRD_3D} (c)]. TTD is reached by a sequence of singular value decompositions, while TRD is realized by recursively solving the self-consistent equations to locate the dominant eigenstates of corresponding matrices. TTD is a local decomposition of the tensor itself, while TRD can be regarded as a global decomposition of the infinite TN that is formed by the local tensor. 

\section{Summary and outlook} 

A simple and fundamental numeric approach named as ab-initio optimization principle (AOP) is proposed to simulate the ground states of translational invariant quantum lattice models with strong correlations. The simulation that contains infinite degrees of freedom is transformed to a local optimization problem in a supercell, where the entanglement between a supercell and the infinite environment are optimally approximated by the boundary states. In AOP, tensor ring decomposition (TRD) is proposed, which is local but encodes the global contraction of the infinite TN.

AOP relies little on the details of the model and has a unified form with TRD. Thus, it is easy to be implemented or commercialized. The current discussions suit spins and bosons. For fermions, the ``entanglement bath'' in AOP should be modified by combining with the density matrix embedding theory \cite{DMET,DMET1,DMET2}.

AOP provides a fundamental picture for existing many-body methods, providing novel connections among the mean-field theory, iTEBD, DMRG, NCD, DMET, rank-1 decomposition and TTD. More properties and applications of AOP as well as TRD in the fields of both many-body physics and multi-linear algebra \cite{TDrev} are to be explored in the future. 

\section*{Acknowledgements} 

The author is indebted to Maciej Lewenstein, Andrew Ferris, Luca Tagliacozzo, Emanuele Tirrito and Gang Su for helpful discussions. The author also acknowledges Sheng-Ying Yue for useful discussions about first principle algorithms, and Garnet Chan for stimulating discussions about DMET. This work was supported by ERC ADG OSYRIS, Spanish MINECO (Severo Ochoa grant SEV-2015-0522), FOQUS (FIS2013-46768), Catalan AGAUR SGR 874, Fundaci\'o Cellex and EU IP SIQS.

\appendix
\renewcommand\thefigure{A}

\section{Ab-initio optimization principle in practice}

For those who intend to use AOP to solve physical problems, a practical introduction of how to implement AOP is given below. This is especially for those who are not familiar with TN algorithms. Steps 1 and 2 are for initialization. Steps 3 to 7 are to calculate TRD. Step 8 is to compute physical quantities.  

Step 1. From the Hamiltonian in Eq. (\ref{eq-H}), choose a supercell and write the shift of bulk Hamiltonian $\tilde{H}^B = \hat{I} - \varepsilon \hat{H}^B/2$ as a matrix in local basis. Do the same thing with the boundary Hamiltonian $\hat{F}^{\partial} = \hat{I} - \varepsilon \hat{H}^{\partial}$. Note that $\hat{H}^B$ contains all interactions inside a supercell, and $\hat{H}^{\partial}$ contains all interactions between two adjacent supercells. The supercell should be chosen so that there is no interaction among non-adjacent supercells.

Step 2. Use singular value decomposition (SVD) to calculate $\hat{F}^L$ and $\hat{F}^R$, as shown in Eq. (\ref{eq-Fboundary}). Calculate the operator $\hat{F}$ as Eq. (\ref{eq-getF}) [also see Fig. \ref{fig-Fboundary} (b)] and restore it as a forth-order tensor. 

Step 3. Give an initial guess of the boundary state $|A \rangle$, which is a third-order tensor. While its elements can be totally random, it is better to be symmetrical for the two ancillary indexes, i.e. $A_{S\mu\mu'} = A_{S\mu'\mu}^{\ast}$, to guarantee the existence of the eigenstates of $\hat{\mathcal{M}}$ in Eq. (\ref{eq-Mmatrix}).

Step 4. Calculate $\hat{\mathcal{F}}$ in Eq. (\ref{eq-Mmatrix}) [also see Fig. \ref{fig-Fminimization} (b)] and its left and right dominant eigenstates $\langle L |$ and $|R\rangle$, and reshape them into third-order tensors. To compute this eigenvalue problem, one can use $\langle L |$ and $|R\rangle$ obtained from the last iteration as the initial guess.

Step 5. Calculate the orthogonal parts of $\langle L |$ and $|R\rangle$ that fulfill Eq. (\ref{eq-isometry}) using SVD [one can see Eqs. (A8) and (A9) in Appendix B] ,and update them.

Step 6. Calculate the effective Hamiltonian $\hat{\mathcal{H}}$ using Eq. (\ref{eq-Heffective}) [also see Fig. \ref{fig-Fminimization} (c)] and its right eigenstate $|A\rangle$. Again, one can use $|A\rangle$ obtained in the last iteration as the initial guess.
	
Step 7. Check if $|A\rangle$ converges. If it does, proceed to Step 8. If not, go back to Step 4.

Step 8. Use the tensor $|A\rangle$ to construct an MPS, which gives the ground state of the system, and calculate the interested physical quantities, such as energy and entanglement. According to the deductions in Appendix B, there is a simple way to calculate observables in AOP. See Appendix C about the computations of observables.

For choosing the value of $\varepsilon$ to shift the Hamiltonian, it depends on the request of precision, since $\varepsilon$ determines the Trotter-Suzuki error that is approximately $O(\varepsilon^2)$. Meanwhile, the smaller $\varepsilon$ is, the more time it will take to do Steps 5 and 6. Luckily, $\hat{\mathcal{H}}$ and $\hat{\mathcal{M}}$ are local matrices, and their dominant eigenvector can be efficiently found by existing eigenvalue algorithms. In practice, a suggested value would be $10^{-3}$ to $10^{-6}$.

\begin{figure}[tbp]
	\includegraphics[angle=0,width=1\linewidth]{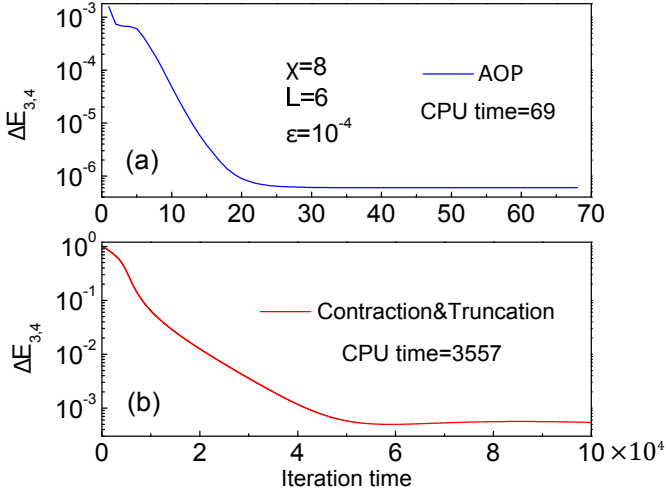}
	\caption{(Color online) The the convergence of the relative error of the bond energy in the middle of the supercell $\Delta E_{3,4}$ along with iteration time. The parameters are $h=0.5$, $N=6$, $\chi=8$ and $\varepsilon=10^{-4}$. One can see in (a) that with the standard AOP, $\Delta E_{3,4}$ converges to $O(10^{-6})$ with only about 30 times of iterations. For comparison, in the contraction-truncation scheme (see the details in the main text in Appendix A), it takes $O(10^{4})$ times of iterations to converge approximately to $O(10^{-4})$. The iterations are stopped when the change of $\Delta E_{3,4}$ after one iteration is smaller than $10^{-10}$. To reach such a convergence, the CPU time that AOP and contraction-truncation scheme takes are $69$ and $3557$, respectively.}
	\label{fig-CPUtime}
\end{figure}

The efficiency of AOP algorithm is shown to be remarkably high. One fundamental progress in AOP is the employment of eigenvalue equations, instead of the contractions and truncations of the TN. Comparing with iTEBD where one only contracts one layer of MPO to the MPS and then truncates (also see Sec. VII), while in AOP, one solves the eigenvalue equations. Fig. \ref{fig-CPUtime} shows the comparison of the efficiency between these two schemes. In Fig. \ref{fig-CPUtime} (a), the standard AOP algorithm is used, where the eigenvalue equations given by Eqs. (\ref{eq-eigen2})-(\ref{eq-eigen1}) are solved in each iteration. In Fig. \ref{fig-CPUtime} (b), the results are calculated in a way similar to iTEBD, i.e. a contraction-truncation way. In each iteration, $L$ and $R$ are updated by solving Eqs. (\ref{eq-eigen2}) and (\ref{eq-eigen3}), which is analog to the canonicalization of MPS aiming at obtaining the environment, and $A$ is updated as $A \leftarrow \hat{\mathcal{H}} A$, which is analog to contracting one layer of MPO with the truncation determined by the canonicalization. The bond energy in the middle of the supercell $\Delta E_{3,4}$ against iteration time is given as an example. Here, I take $h=0.5$ (critical point), $\chi=8$, $N=6$ and $\varepsilon=10^{-4}$. One can see in Fig. \ref{fig-CPUtime} (a) that with AOP, it only takes about 30 times of iterations for $\Delta E_{3,4}$ to reach $O(10^{-6})$. For the contraction-truncation scheme in Fig. \ref{fig-CPUtime} (b), it takes $10^4$ times of iterations to converge to $O(10^{-4})$. For the total CPU time, AOP is faster almost by two orders of magnitude.

\section{Constraints in ab-initio optimization principle}

The purpose of this section is to prove that the local eigenvalue equation that $A$ satisfies is the one given in Sec. IV, i.e., to prove why $\tilde{L}$ and $\tilde{R}$ can be chosen as the optimal isometries obtained from the SVD of $L$ and $R$. Meanwhile, the deduction below will show how the entanglement spectrum naturally appears in AOP. I will use graphic representations of the equations to present.

From the algorithm shown in Sec. IV (also in Appendix A), one can see that $\langle L|$ and $|R\rangle$ satisfies 

\begin{center} \includegraphics[angle=0,width=1\linewidth]{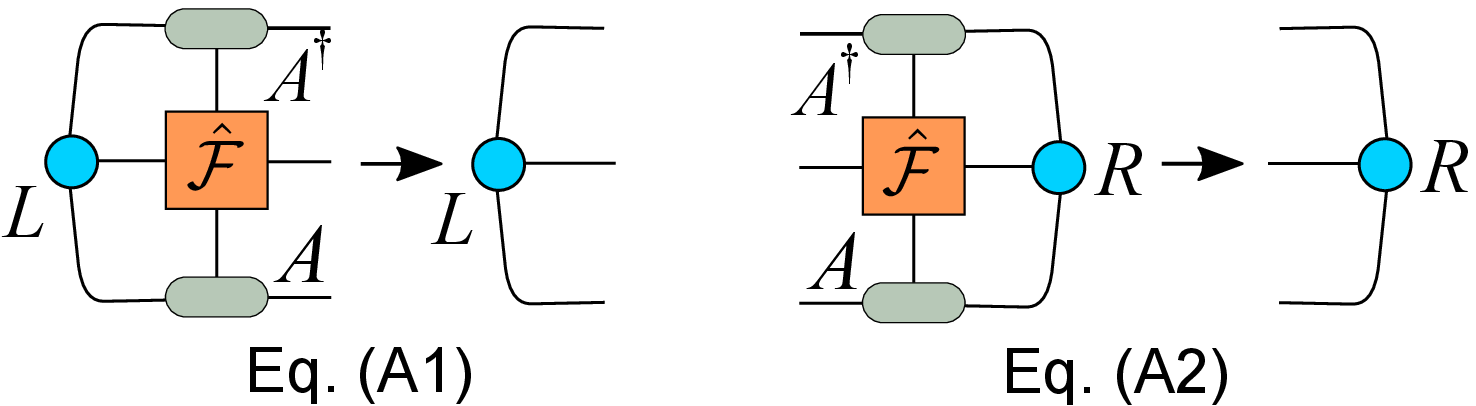} \end{center}

The arrow means after contracting all shared bond on the left-hand-side, the result is given by the object on the right-hand-side multiplied by a constant. Eqs. (A1) and (A2) are fulfilled under the constraint Eq. (\ref{eq-constraint1}) while optimizing the function Eq. (\ref{eq-Ffunction}). 

From the constraint given by Eq. (\ref{eq-constraint2}), it is easy to find that the optimization of the tensor $A$ results in a generalized eigenvalue problem

\begin{center} \includegraphics[width=2.8in]{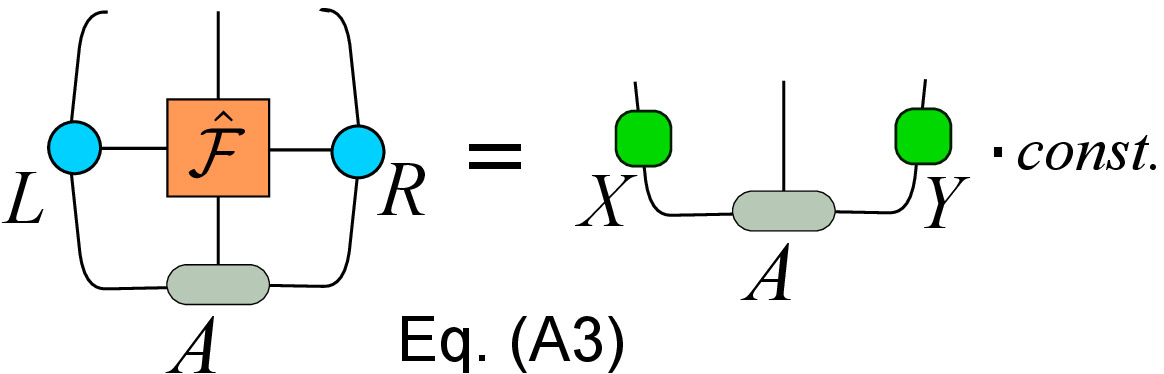} \end{center}

For simplicity, we set the constant in Eq. (A3) as 1. $X$ and $Y$ are the left and right dominant eigenvector of the matrix formed by $A$ and its conjugate, which read

\begin{center} \includegraphics[width=2.9in]{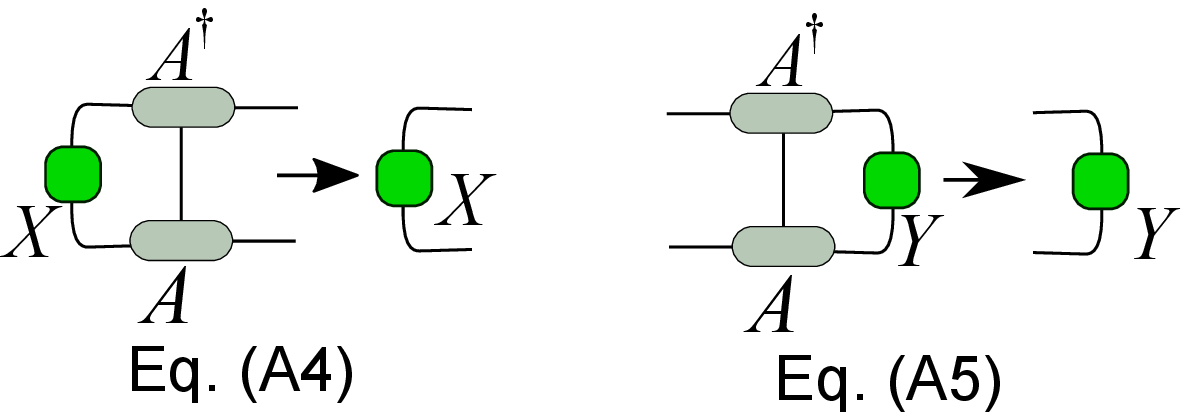} \end{center}

Normally, $X$ and $Y$ are Hermitian (because of the structure of the matrix formed by $A$ and $A^{\dagger}$). It is not convenient and easy to directly solve Eq. (A3). One possible problem is that $X$ and $Y$ might be singular, and the generalized eigenvalue problem cannot be transformed into a regular one by simply taking their inverses. Here, we suppose that the local (regular) eigenvalue equation of $A$ with the constraint Eq. (\ref{eq-constraint2}) fulfilled is given by $\tilde{L}$ and $\tilde{R}$, which reads

\begin{center} \includegraphics[width=1.7in]{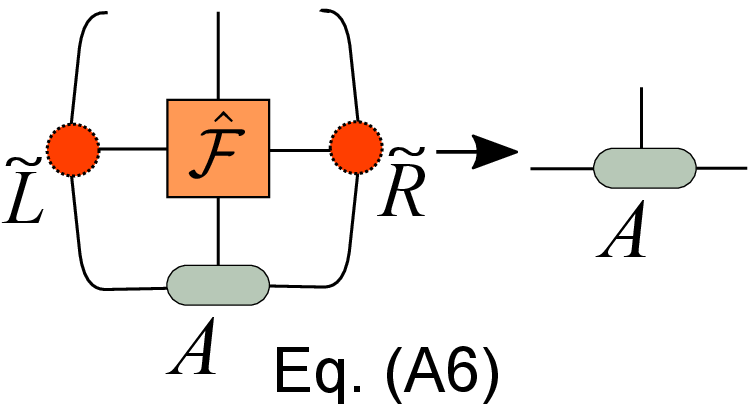} \end{center}

One needs to find out what $\tilde{L}$ and $\tilde{R}$ are. To this end, substitute Eq. (A6) into the right-hand-side of Eq. (A3)

\begin{center} \includegraphics[width=3.5in]{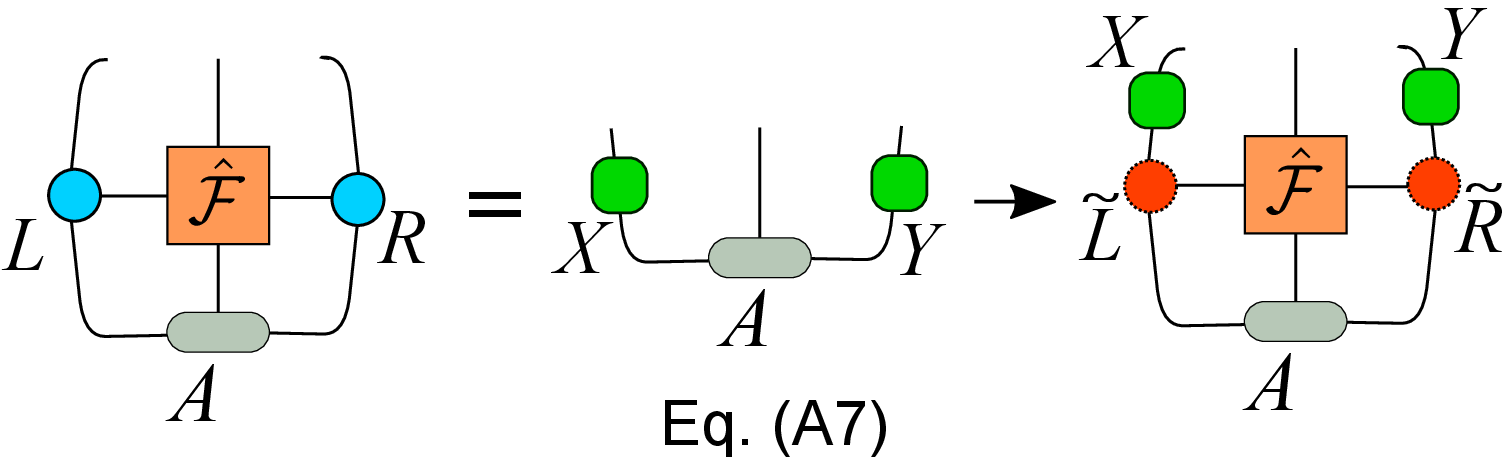} \end{center}

Compare the first and last expressions in Eq. (A7), one has

\begin{center} \includegraphics[width=2.6in]{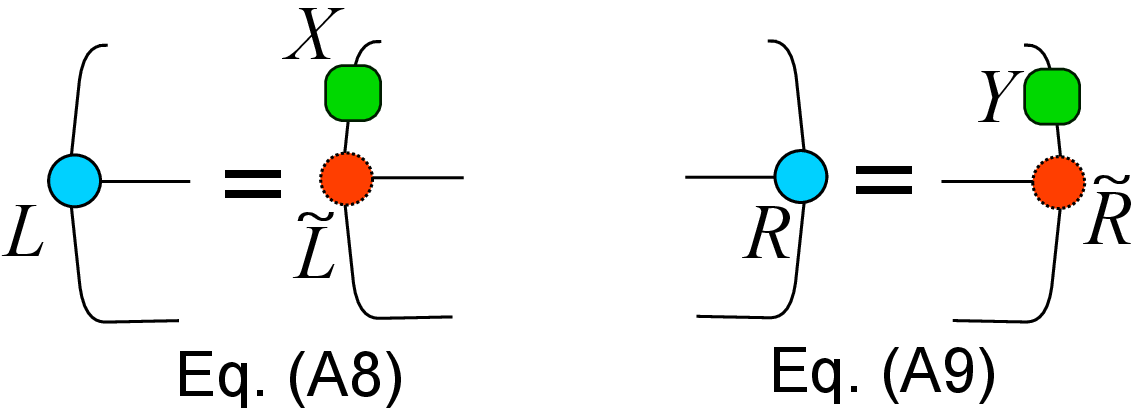} \end{center}

Note that any constants appearing on the right-hand-side of Eqs. (A8) or (A9) can be absorbed into $X$ or $Y$. Substituting Eqs. (A8) and (A9) into Eqs. (A1) and (A2), respectively, one has

\begin{center} \includegraphics[width=3.1in]{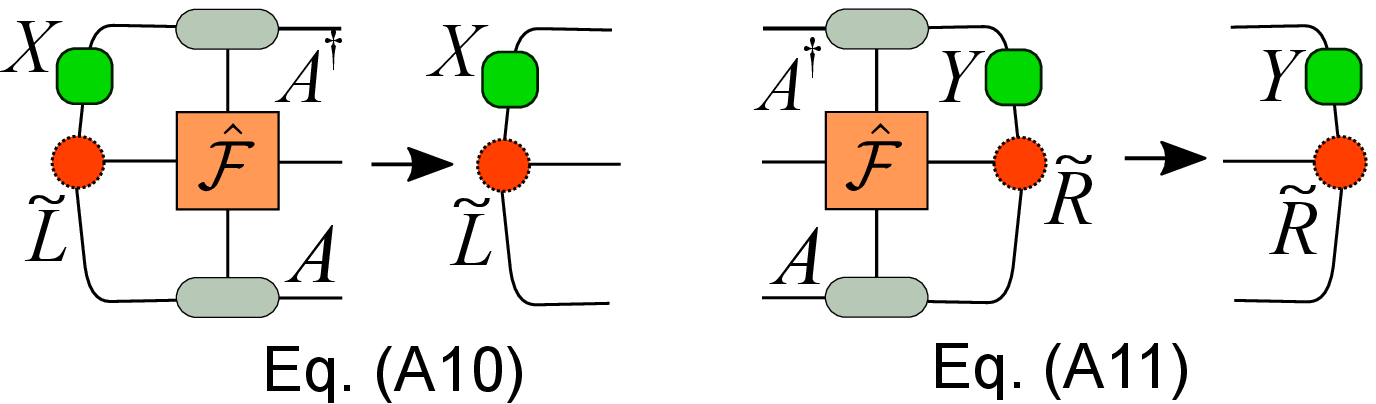} \end{center}

By multiplying $\tilde{R}$ on both sides of Eq. (A10), and multiplying $\tilde{L}$ on both sides of Eq. (A11), one uses Eq. (A6) again and obtains

\begin{center} \includegraphics[width=3.0in]{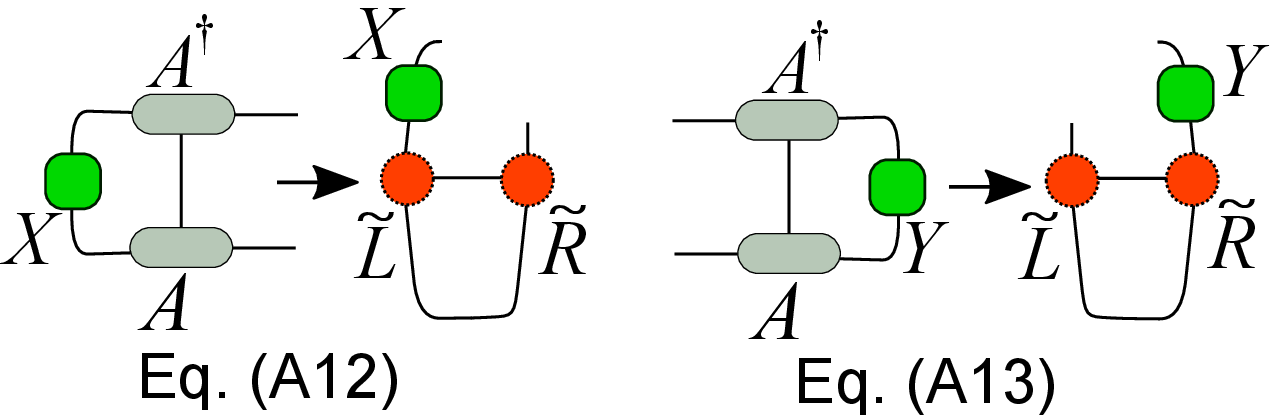} \end{center}

By comparing Eqs. (A12) and (A13) with the eigenvalue equations of $X$ and $Y$ given by (A4) and (A5), one has the restriction for $\tilde{L}$ and $\tilde{R}$, which is

\begin{center} \includegraphics[width=2.1in]{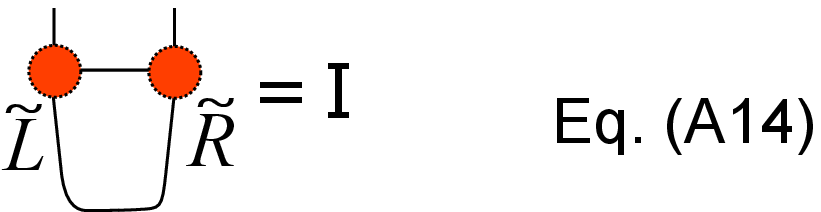} \end{center}

Under the assumption that $\tilde{L}$ and $\tilde{R}$ are conjugate to each other, Eq. (A14) directly leads to the orthogonal conditions given by Eq. (\ref{eq-isometry}) in Sec. IV.

With the knowledge that $\tilde{L}$ and $\tilde{R}$ are orthogonal isometries, one can readily see that Eqs. (A8) and (A9) can be reached by singular value decomposition (SVD) of $L$ and $R$ that read
\begin{eqnarray}
 L=USV^{\dagger}, ~ R=\mathcal{U}\mathcal{S}\mathcal{V}^{\dagger}. \nonumber
\end{eqnarray}
Then, one has
\begin{eqnarray}
  \tilde{L}=UV^{\dagger}, ~ L=X\tilde{L}, ~ X=USU^{\dagger}, \nonumber \\
  \tilde{R}=\mathcal{U}\mathcal{V}^{\dagger}, ~ R=Y\tilde{R}, ~ Y=\mathcal{U}\mathcal{S}\mathcal{U}^{\dagger}. \nonumber 
\end{eqnarray}
It can be also seen that since $S$ (note $\mathcal{S}=S$) is singular spectrum that is real, $X$ and $Y$ are naturally Hermitian, which is consistent with the AOP theory. In fact, $S$ is the entanglement spectrum of the ground state. This is because $S$ is the eigenvalue spectrum of $X$ and $Y$, which are the left and right dominant eigenstate of the transfer matrix that forms $\langle MPS| MPS \rangle$ (with $| MPS \rangle$ the ground state). Thus, $X$ and $Y$ represent the reduced density matrix (in the ancillary space) of the left and right infinite halves of the system, respectively. The entanglement spectrum $S$ is obtained by diagonalizing such reduced density matrices with $U$ (or $\mathcal{U}$).

\section{Calculations of observables}

It is very easy to calculate observables in AOP. Knowing that the ground state $| \Phi \rangle$ is actually an MPS, the observable $\langle \hat{O} \rangle$ becomes the contraction of the operator, the ground state MPS and its conjugate, i.e. $\langle \hat{O} \rangle = \langle \Phi|\hat{O}| \Phi \rangle/Z$ with the normalizing factor $Z = \langle \Phi| \Phi \rangle$. In AOP, one in fact does not have to calculate the whole contraction. From the deductions in Appendix B, we know that with the SVD of $L$ and $R$, one already has the dominant left and right eigenstates of the transfer matrix of $\langle \Phi |\Phi \rangle$, as shown in Eqs. (A4) and (A5). Taking the supercell size $N=4$ as an example, the bond energy in the middle of the supercell $E_{2,3}$ and that between two adjacent supercells $E_{0,1}$ can be written as

\begin{center} \includegraphics[width=3.4in]{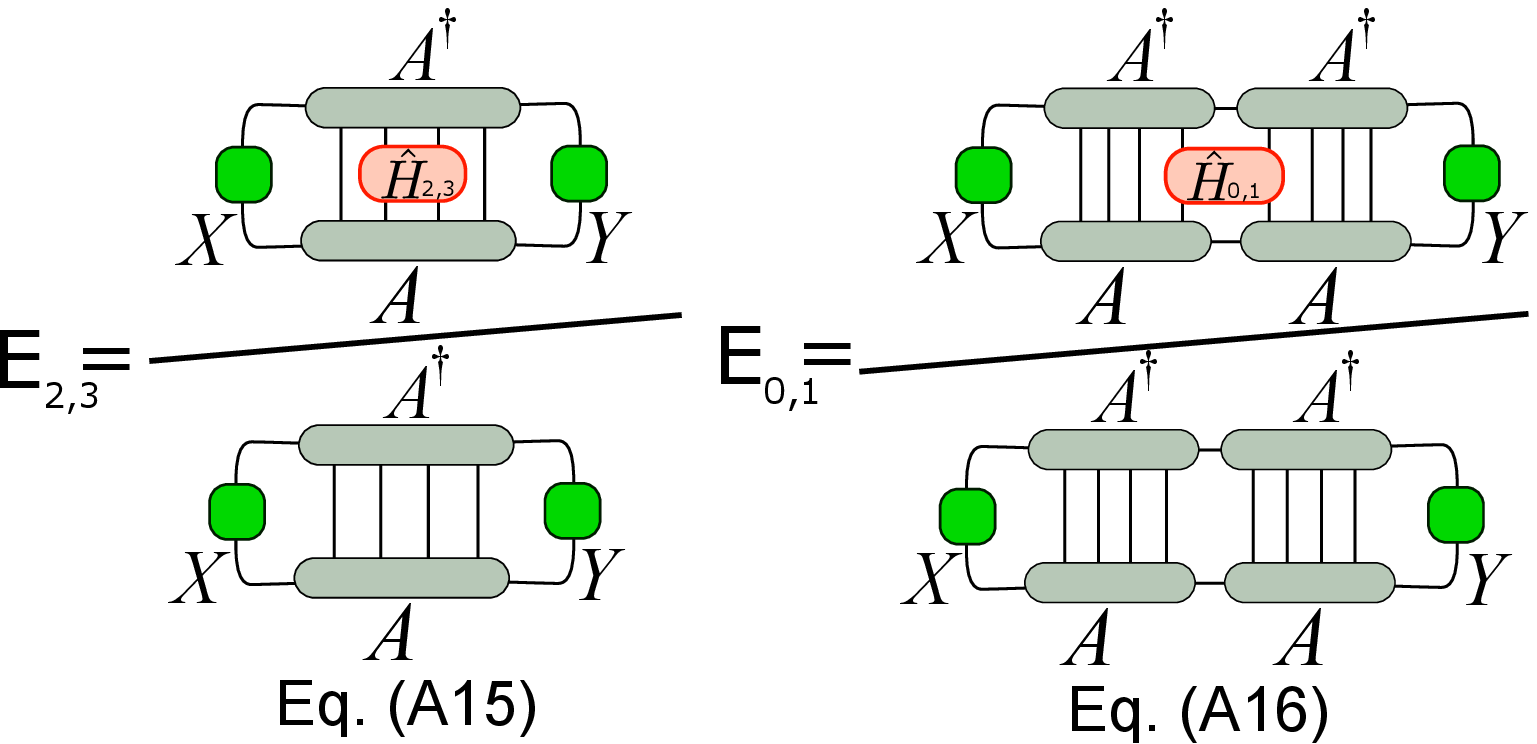} \end{center}

Eqs. (A15) and (A16) can be easily generalized to other observables.

\end{document}